# Secure Software Development: Issues and Challenges

**Sam Wen Ping, Jeffrey Cheok Jun Wah, Lee Wen Jie, Jeremy Bong Yong Han, Saira Muzafar**

*School of Computer Science, SCS,* **Taylor's** *University,* **Subang Jaya Malaysia**
wenping.sam@sd.taylors.edu.my, jeffreycheokjunwah@sd.taylors.edu.my,
leewenjie@sd.taylors.edu.my, jeremyyonghan.bong@sd.taylors.edu.my,
sairamuzafar@hotmail.com

**Abstract:** *In recent years, technology has advanced considerably with the introduction of many systems including advanced robotics, big data analytics, cloud computing, machine learning and many more. The opportunities to exploit the yet to come security that comes with these systems are going toe to toe with new releases of security protocols to combat this exploitation to provide a secure system. The digitization of our lives proves to solve our human problems as well as improve quality of life but because it is digitalized, information and technology could be misused for other malicious gains. Hackers aim to steal the data of innocent people to use it for other causes such as identity fraud, scams and many more. This issue can be corrected during the software development life cycle, integrating security across the development phases, and testing of the software is done early to reduce the number of vulnerabilities that might or might not heavily impact an organisation depending on the range of the attack. The goal of a secured system software is to prevent such exploitations from ever happening by conducting a system life cycle where through planning and testing is done to maximise security while maintaining functionality of the system. In this paper, we are going to discuss the recent trends in security for system development as well as our predictions and suggestions to improve the current security practices in this industry.*

***Keywords: Software Security, Secure Software Development, organizational factors, SSDLC, Secure Software Development Life Cycle***

# 1. Introduction

The exponential expansion of the internet has brought about a transformative shift in business processes. Most of public and private enterprises conduct their daily activities by utilizing web applications [1]. The inadequately designed software systems might give rise to vulnerabilities that can be manipulated for illicit uses, violating software security principles. Software security breaches have become increasingly prevalent in recent times, with a significant proportion attributed to flaws in software architecture. Considering the increasing reliance of individuals and businesses on software systems for their daily operations, it is imperative to prioritize the development of safe software solutions. The software development normally follows two popular models that are the Waterfall and Agile models. The waterfall method is very linear and follows phases in which are very orderly and tend to avoid any risky designs. There are many disadvantages to a Waterfall modelled project as there is no indication where the clients may want any alterations after seeing the final product of the software development. Thus, if there were any requests to change the software, it would incur increased costs and the whole process of software development would be restarted. The other model is the Agile model, it uses fast paced methodology such as extreme programming and scrum, it highly implements prototyping along the software development cycle. This includes giving weekly or monthly demonstrations of the software to stakeholders. The result of this Agile model is a self-reinforcing model.

A secure software development cycle was introduced after realising that just having a well-structured software development cycle isn't going to help improve security and reduce vulnerabilities of the software developed. So, to combat this, a new and common practice to the software development industry is SSDLC (Secure Software Development Life Cycle). This practice involves running security-based activities only as a part of testing at every phase of the software development process, this is to prevent or reduce the number of issues and vulnerabilities early and in the midst incorporating a security aspect in every detail of the SDLC. Although the goal of the SSDLC is to improve the security of the product, this is only at the cost of the performance of the product, the integrated security functions must not be a nuisance or bother the performance of the product. There are many standards of SSDLCs that have been proposed namely, the Microsoft Security Development Lifecycle (MS SDL), NIST 800-64, and OWASP CLASP [2].

Because the SSDLC model is a model where various standards are used to achieve the best practices and results with security. Not all models have the same instructions. The general trend for secure software development comprises 5 phases: requirements engineering, system architecture, development, execution and testing, and quality management as shown in Figure 1.

a. Requirements Engineering

The first phase of SSDL is requirements engineering, this phase introduces the purpose of the software and its environment to analyse and make clear the exact requirements including text, features and traceability for the requirements. Overall, in this phase the team needs to determine the scope of the project and the desired security level and to identify security requirements. Analysis of potential risks and threats is also required. The requirement gathering process involves getting the approval of stakeholders and peers on how the software is going to be developed and how it would be configured.

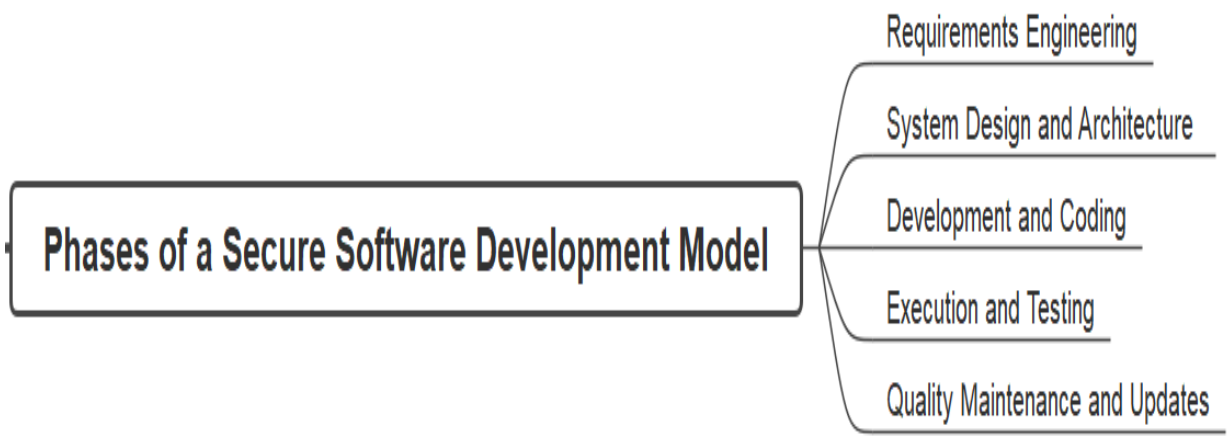

Figure 1: Phases of secure software development model

b. System Design & Architecture

Next to investigate and define a secure architectural design with appropriate security control and mechanism. Create a detailed design and architecture of the software, incorporating security features and determine the best security controls and mechanisms to be implemented. Planning the system architecture allows the developers to identify core components, modules, interfaces, and information for a software system to fulfil its general purpose based on the Method Task Establish and Maintain Software Architecture [3].

### c. Development and Coding

During the development phase write and code the software according to the design and security specifications. The current trend is to follow all the standards for the secure coding practices to avoid SQL injections, XSS, DoS and DDoS which disrupt and damage the network activates [4]. The coding should be continuously reviewed for security vulnerabilities and built around its vulnerabilities and validate its build for the entire system [5].

### d. Testing and Execution

Following that, execution and testing will be done to validate the behaviour of the software as well as its characteristics. This phase allows developers to identify any bugs and relinquish any unwanted functions from the software. The decorum of testing phase is to conduct thorough security testing to identify any vulnerabilities or weaknesses in the software and to verify compliance with security requirements. The team can use both automated tools and manual techniques to detect potential security flaws. For the execution deploy the software in a secure environment, following best practices and to ensure that all relevant security measures, such as access controls and encryption, are properly implemented. Release only after thorough security assessment.

### e. Quality Maintenance and Updates

Lastly, the quality management phase aims to provide an assurance that the product or software released will fulfil its functional and non-functional requirements and that everything in the software will meet the standards.

The current SSDLC can be summarised with reviewing code in every step of the development phase. The reviewing of code is so important that it allows the identification of early problems as well as risks that may be imposed later. Thus, reviewing code in every phase of the SSDLC is important and most popular practices do that as well. The challenge in this task is that a software's code cannot be reviewed entirely as it would take too much time and the block of code would just be too massive and long for a human to slowly scan through it even with the help of debuggers and vulnerability scanners, thus, a review in each incremental addition of code and each development phase would be optimal. Further, it is extremely important to regularly update and patch software to address any newly identified vulnerabilities. Monitor for security incidents and respond promptly to mitigate risks.

The rest of the paper is organized as follows: Section 2 presents the issues and challenges in secure software development 3 presents the discussion, and Section 4 concludes the paper.

## 2. Issues and Challenges in Secure Software Development

To be able to provide a Secure Software, the 'culprit' that hindering its performance must first be identified. An In-depth Literature Review (ILR) has been carried out and few issues and challenges have been categorised as weakening or causing an unsecure software. These are Organisational Factor, Agile Approach in Secure Software, Lack of Resources and Law and Regulation towards Software Development.

### 2.1. Issue within Organisational Factor - *Lack of Policy Enforcement*

Recently, a qualitative analysis towards Malaysian software has been published by the Indian Journal of Science and Technology indicating that the Organisational Factor has been the reason for weakening the Secure Software [3]. The lack of policy enforcement such as no proper guideline showing the lack of concern towards developing a Secure Software from the organisation. This linked to the organisation's belief or perception about whether the product would be the target of an attacker or hacker, which is conflict as mostly it is opined by top management who do not have background and knowledge for a Secure System. For instance, a B2B application has been created and with the belief by the organisation that it would not be the target of an attacker turns out, the large growing user base is attractive and prompt to be attacked by the attacker. Some data that is sensitive including finance, health or education related content, in other words data that users are personally identifiable.

### 2.2 Issue within Organisational Factor - *Lack of Motivation*

The motivating culture embraced in the organisation does not inspire software engineers or developers to constantly develop a Secure Software. This is closely related to the profit maximisation motive of the organisation. It would only provide incentives such as allowance, bonus, gifts and certificates to projects or products that achieve huge profit in return, and worse, mostly do not have background in software, and do not have security awareness in software. Software engineers or developers would be unlikely to spend time and effort to learn and investigate how to make the software more secure, as there is a performance expectation standard set by the organisation, fast and efficient. They would not want to delay their performance in finding out the weakness in software as organisations only want the selling points which are the advantages of the software rather than the throwing points. The lack of training provided by the organisation has also been the disseminating factor resulting in developer lack of intention and even security awareness towards the Secure Software. The training is supposed to increase engineers and developer's skills and learn the current software technologies and methodologies [6]. As the organisation's motive is in profit maximisation, they would not be willing to spend thousands or even hundred thousand on facilities for developer research on secure development. Interviews carried out in the qualitative analysis showed that developers are not satisfied with the available facilities condition and expressed that the security development research would have been better if they had proper testing tools and teams.

## 2.3. Issue within Agile Approach to Secure Software

The Agile Approach begins with seventeen software developers in Wasatch Mountains of Utah, USA discovering the manifestos and principles that focus on flexibility in software development, the working software, and customer's needs. Figure 2 and 3 shows the manifesto and principles that were discovered by the developers.

> We are uncovering better ways of developing
> software by doing it and helping others do it.
> Through this work we have come to value:
>
> **Individuals and interactions** over processes and tools (V1)
> **Working software** over comprehensive documentation (V2)
> **Customer collaboration** over contract negotiation (V3)
> **Responding to change** over following a plan (V4)
>
> That is, while there is value in the items on
> the right, we value the items on the left more.

Figure 2: Manifesto for Agile software development

P1. Our highest priority is to satisfy the customer through early and continuous delivery of valuable software.
P2. Welcome changing requirements, even late in development. Agile processes harness change for the customers competitive advantage.
P3. Deliver working software frequently, from a couple of weeks to a couple of months, with a preference to the shorter timescale.
P4. Business people and developers must work together daily throughout the project.
P5. Build projects around motivated individuals. Give them the environment and support they need, and trust them to get the job done.
P6. The most efficient and effective method of conveying information to and within a development team is face-to-face conversation.
P7. Working software is the primary measure of progress.
P8. Agile processes promote sustainable development. The sponsors, developers, and users should be able to maintain a constant pace indefinitely.
P9. Continuous attention to technical excellence and good design enhances agility.
P10. Simplicity—the art of maximizing the amount of work not done is essential.
P11 The best architectures, requirements, and designs emerge from self-organizing teams.
P12. At regular intervals, the team reflects on how to become more effective, then tunes and adjusts its behavior accordingly.

Figure 3: Code Principle for Agile software development

**2.4 Issue within Agile Approach to Secure Software -** *Security Assurance*

As shown in the figures 2 & 3, it focuses more on working of the software rather than the comprehensive documentation. It means that this approach encourages minimum work and evaluates the progress based on working of the software [7]. It utilises light documentation that conflicts with security assurance such as in System Security Engineering-Capability Maturity

Model (SSE-CMM) framework that requires detailed documentation for Secured Software developers research in ensuring the security as shown in Figure 4. Some detail issues within the security might be overlooked and further neglected by the security evaluators in this approach. The report is too general and unable to fulfil the security requirement [8]. Even automation of the documentation tools is hard to achieve detailed reports as customization of format is needed.

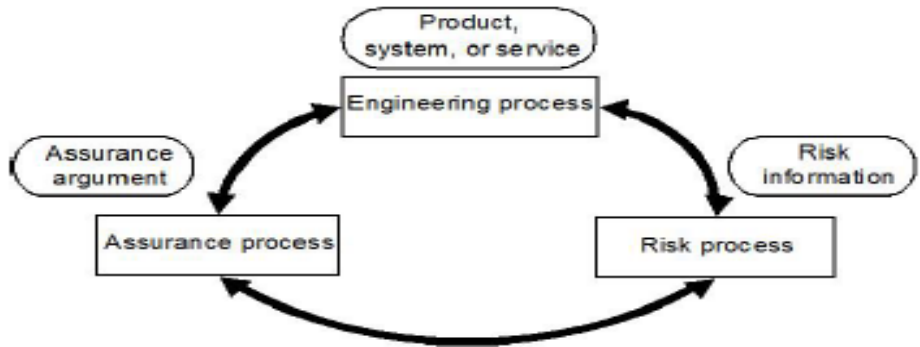

Figure 4: CMM Security Engineering Process

This is also in reliance on their main principle that embraces responding to customer needs and changes requested. Whenever there is a request from a customer, developers need to adjust or changes pursuant to that request. Thus, code refactoring is a common practice in the Agile Approach. Note that frequent code changes make the security assurance activity difficult because the changes may result in security reviews, and tests invalidated [8]. It made security requirement tracing even more challenging and difficult. The method of assessing use by the security evaluators in Agile Approach does not match the requirement as well. This is because security evaluators use development process-related information such as architecture documents in assessing the software security. This method is inaccurate especially when the development process has been changed due to customer requests [9].

**2.5. Issue within Agile Approach to Secure Software - *Cost and Profit Prioritisation***

Many of the organisations tend to compromise the security level of the software especially when it reaches the releasing schedule. As mentioned before, organisations are in the motive of profit maximisation, they rather spend the time and cost for innovation that is attractive to its customer. Developers on the other hand are mainly responding to customer needs. Customers that have low security level awareness would not be able to realise that their personal data in the software needed stronger security. The organisation would also not state the software weaknesses to the customer as they are selling the product [10]. All these result in weak security levels within the software.

**2.6. Issue due to Lack of Resources and Facilities**

The issue of lack of resources typically seen in start-up organisations that have not enough models to set up complete facilities [11]. For instance, the incomplete vulnerable scanning solution. This causes some misconfiguration and vulnerabilities unable to be detected. Most organisations often provide one solution or method in assessing the information, with the opinion that 'one size fits all' is not very helpful in providing secured software. The result is inaccurate and unable to use in assessing the exact error or missing patches that have existed in the software. The poor-quality testing in assessing the security of the software might result in the system being more prompt to attackers and vulnerabilities as well [12]. This is due to delay and slowness in the manual patching system and difficulty to replicate the exact production state during testing due to lack of technical skills and knowledge.

There is also a missing central platform to patch the information received for retrieval and filtering in the Software Security Patch Management System [13]. The lack of automation in retrieving the information causes more human resources to be needed, with the probability of overlooking and human error might have occurred. Some customised software requiring expertise may be unable to be hired by the organisation as it is costly. This is because professional security training and security certification are often seen as too costly to implement in terms of time and resources while their value is questioned by many organisations that are profit motive [14].

**2.7. Issue due to Law and Regulation**

The issue within Law and Regulation can be seen in government policies. The regulations are unclear and quickly outdated with the advancement of technology. This is often seen in the conflict between government sovereignty over privacy which is the main ingredient that the security is taking care of. The government claims that they have sovereignty over the privacy of individuals that can be entangled with national interest. For instance, organisations cannot refuse to provide personal data to the government in line with national interest, the so-called, 'backdoor' [15].

The laws are often vague and require subjective human judgement and expertise to interpret the law in practice. This is where the Siloed Team becomes apparent, where the legal team and software developer team are separated in general [16]. The lack of collaboration between the teams could not be blamed as lawyers and developers do not speak each other's languages. For instance, the jargon and conceptual framework that learn and understand are totally different and this creates communication issues at different levels. Often, privacy professionals locked in their legal compliance preventing the access of the developers' expertise [17]. The lack of diversity of the team from different backgrounds causes a certain perception that they thought is useful becomes impractical to perform. For instance, developers think that certain privacy mechanisms can be easily broken and overridden while the guidelines for implementing such mechanisms are complex and too theoretical to be used in practice [18]. For lawyers, due to the lack of in-depth knowledge in the software, they might think that the law is sufficient to protect the user; however, hackers could find loopholes in avoiding legal compliance.

### 2.8. Issue within Security of Open Source Software

Open-Source Software (OSS) is a software that is open to public uses, consist of source code as is of available for modifications, and it typically has no charges [19]. It must meet the criteria of Redistribute the software without restriction; modify computer programs, Access and modify the source code, Distribute the modified version of the software. For instance, Linux and Python programming languages.

Some developers who perform malpractices from its work tend to copy and paste the code from the open-source libraries [20]. This is problematic as it will create vulnerability from the copied piece of code. The OSS vulnerabilities are made available for everyone to view on the National Vulnerability Database (NVD) [17]. Attackers may easily exploit and break the code via the database if developer malpractice has been noticed. There is also no way to track and update a code snippet once it is added into the codebase[21]. Furthermore, some organisations have the possibility of failing to update the open-source component when a newer version is available [22]. This will cause catastrophe as the case in Equifax breach.

### 3. In-depth Discussion

Based on the literature review that has been discussed earlier, challenges and issues are progressively occurring within the secure software development process as nowadays almost everything around us is virtually evolving into technology [23]. As mentioned earlier, it is difficult to ensure and track the quick expansion of an attack's hacking and threat capabilities. The reason for this is that as security evolves, new problems appear at random, offering attackers the chance to attack if there are flaws in the secure software development phases. The issues and challenges are classified into three levels which are less critical, medium, and critical.

There are 2 issues and challenges that are deemed essential due to using the same technique. One of the first is an organisational factor, which due to lack of motivation, causes organisations to prioritise profit over producing safe and dependable software **[23].** As a result of the lack of trustworthy security awareness in the programme, numerous forms of errors and flaws arise throughout the development process [24]**.** The second significant issue is the issue of cost and profit priority demonstrated by the agile strategy. Their primary purpose is to maximise profits, which leads to a disregard of software development [25]**.** Customers are unaware that their credential information has been exposed because of their involvement because all they know is that their needs have been granted. Both have been regarded as major concerns due to their similarity. Software security advances are not being built and designed appropriately; users' credentials will be disclosed. Not only that but having a short-term goal will have a big influence. As a result of reckless and unreliable software security development that involves several gaps, customers will be notified that their credential information has been disclosed. With this, the numbers of cyber-attacks towards software development increase significantly.

Moving on, another parallel discovery which considers as another critical issue or challenge that appear between the organisation and the agile method. This issue and challenge arise because of inadequate legislation and policy enforcement, which raises concerns throughout the system software development process [26]**.** As previously said, system software development is divided into five phases, each must be controlled and work effectively. For starters, having ambiguous and out-of-date instructions will lead to confusion. As a result, confusion doubles and will emerge on a frequent basis during the development phases process, as everyone has a different attitude about it. This will lead to processing the requirements thus wasting time and effort. Furthermore, since the number of users grows day by day, certain organisations and businesses violate the law [27]**.** They will tend to or be oblivious of allowing additional users even if they are at full capability and incapable of handling. Due to the security measures' inability to meet the need, this would result in frequent breaches as flaws surface during the development phases. As some users are dissatisfied with the law enforcement, which leads to poor collaboration across the development team owing to a lack of diversity may ensue. Besides, due to the knowledge included in the system software development, various needs and goals cannot be achieved due to the overcomplexity. As a result, security dangers such as identity theft, hacking and others emerge.

Moreover, because of certain similarities with the lack of resource and facilities issue, the security assurance issue that happened in the agile method will function in the system software development is classified as a medium type of issue which has less impact [28]. This occurred because of taking shortcuts in completing the task by ignoring certain critical areas of the development. In brief, finishing the workload faster to reach the goals rather than following the workload faster to reach the goals rather than following the standards, which involves more effort and time. Some security precautions may be neglected, resulting in catastrophic repercussions. For example, cyber-attacks or even demolishing the development and restarting it. Furthermore, as the customer's demands change, considerably more work is required since their goals do not satisfy the customer's expectations because of taking shortcuts [29].

Another issue and challenge that happened and is classified as a medium type of issue which is due to lack of resources and facilities. The cause of this problem is due to lack of capability and requirements to meet its objectives and expectations. This was a common occurrence in medium and small-sized organisations and businesses. As technology is evolving quicker than ever, several medium and small-sized businesses are unable to obtain the most up-to-date technology due to limited funds and resources. With this, certain vulnerabilities and threats will go undetected, resulting in major repercussions such as reputational harm or even bankruptcy. Another factor that contributed to this was the failure to acquire qualified managerial competence. Due to the lack of competence or professional management of the security of software system development, attackers may execute cyber assaults on a regular basis [30]. As threats increase rapidly nowadays, such major and serious vulnerabilities are unable to be solved without the experience of specialists [31]. As a result, user data is readily leaked thus harming the company's reputation due to not being capable of keeping user's data safe and secure.

Finally, there are lots of debates about the agreement and disagreement regarding the issues and challenges that arise. However, we as users must take responsibility for playing our part in ensuring that security is prioritised. This ensures that our credentials are kept safe and secure from unauthorised users.

## 3.1. Proposed Solution

We have proposed several solutions for some of the challenges identified earlier. Firstly, we recommend that companies and organisations should create and enforce an IT security policy that should be adopted by all departments and employees. This policy should contain clear and well-defined information regarding security. This includes what security practices should be implemented in the company's software-based products as well as what steps should be taken in case of a security issue, such as an attack. The company should also assign staff at various departments and levels across the company who will ensure that the policy is being enforced equally by all staff. It is also advised that companies create high-level positions particularly for people with knowledge in secure software development because top management rarely consists of employees who specialise in this field.

Besides that, companies should also offer bounties for penetration testing for their software. Ethical hacking or penetration testing is a program where companies invite ethical or white-hat hackers to find and exploit vulnerabilities in their systems in exchange for a payment [32]. Several well-known companies in the IT industry already have formal programs like this, such as Microsoft who offers up to US$15000 for each vulnerability discovered. An actual malicious attack on a company's software could result in large losses for the company, even more than the profit they would have gained from releasing the software early without implementing essential security features. New vulnerabilities and attacking techniques are being developed every day, and knowing the potential impact of insecure software would encourage the top management to take it more seriously.

In most countries, there are security and privacy-related laws which digital software and applications must comply with. As a whole, more collaboration is needed between people in the IT and legal industries in both companies and governments side. For companies, this is to ensure that their products are compliant to local regulations, as well as foreign laws if the products are available for consumers worldwide [33]. Legal experts should be integrated into software development teams, and IT and legal experts should communicate regularly at every stage of development to discuss details of the software to resolve any issues**.** For example, if new laws are created, the legal team should inform developers on aspects of the software which need to be changed and the deadline before the law goes into effect.

Additionally, this collaboration between IT and legal industry experts should also apply to the government and lawmakers. When new laws related to IT and software security are created, it is better if the lawmakers who draft the bill personally have a background in software development or can consult with a person who does [34]**.** This is to ensure that the laws can be clearly understood and are realistic.

In [35-40], the issues related to secure software development can impact operations for all the machines at all levels, such as any operations related to wireless devices and their various types of operations, including routing, detection capabilities, etc. This lack of software processes can impact hardware performance, which is supposed to run the resulting software [41-48]. Further, the tiny software used for smart cities, smart gadgets, IoT devices, IoT devices performance and security gets affected [49-58] due to the quality compromises of the secure software development process. Ransomware and other related attacks can easily be launched and compromised [59-61] due to this lack of proper following secure software development approach.

## 4. Conclusion

Nowadays, nearly everything around us is virtually evolving into technology, issues and challenges are increasingly occurring and have become more common within secure software development. With technology constantly advancing, the necessity for security has become a major topic of discussion. It is not easy to address the issues and challenges since correct standards, requirements and measures need to be carried out. A consideration about implementing and carrying out the waterfall and agile models concurrently to achieve the best results and outcomes.

Multiple reviews and guidelines are necessary throughout the 5 secure software development checkups and guidelines are required to verify that each software system development step is carried out as planned. Assuring that any misunderstanding is identified and addressed in the correct manner. Not only that but ensuring that all members of the system secure software development share the same mentality in attaining their aims.

If an issue or challenge is discovered, notifying, and labelling of the issue must be carried out immediately. It has been demonstrated that addressing the issue early can result in cheaper maintenance costs than addressing it later, which can result in catastrophic effects. Not only that but making sure that the software within the development is kept up to data is also crucial. Threats may emerge if secure software advancements are unable to be maintained.

Based on the ongoing system development security debate, the predictions and proposals made to enhance security processes should be given top priority. Having secure and safe security measures included as part of the software system development process helps lessen the chance of difficulties and obstacles occurring. Everyone plays a role in ensuring that effective security measures are carried out as planned, since serious consequences will occur if it is not being carried out properly as intended.

**Conflict of Interest:**
Authors declared we don't have any conflict of interest.

**References**

[1] S. Muzafar and N. Z. Jhanjhi, "Success Stories of ICT Implementation in Saudi Arabia," *https://services.igi-global.com/resolvedoi/resolve.aspx?doi=10.4018/978-1-7998-1851-9.ch008*, pp. 151–163, Jan. 1AD, doi: 10.4018/978-1-7998-1851-9.CH008.

[2] N. Drucker and S. Gueron, "Fast modular squaring with AVX512IFMA," *Adv. Intell. Syst. Comput.*, vol. 800 Part F1, pp. 3–8, 2019, doi: 10.1007/978-3-030-14070-0_1/TABLES/4.

[3] M. A. Hamid, Y. Hafeez, B. Hamid, M. Humayun, and N. Z. Jhanjhi, "Towards an effective approach for architectural knowledge management considering global software development," *Int. J. Grid Util. Comput.*, vol. 11, no. 6, pp. 780–791, 2020, doi: 10.1504/IJGUC.2020.110908.

[4] S. Muzafar and N. Jhanjhi, "DDoS Attacks on Software Defined Network: Challenges and Issues," *2022 Int. Conf. Bus. Anal. Technol. Secur. ICBATS 2022*, vol. 2022-Janua, 2022, doi: 10.1109/ICBATS54253.2022.9780662.

[5] M. Humayun, M. Niazi, N. Z. Jhanjhi, S. Mahmood, and M. Alshayeb, "Toward a readiness model for secure software coding," *Softw. Pract. Exp.*, vol. 53, no. 4, pp. 1013–1035, Apr. 2023, doi: 10.1002/SPE.3175.

[6] K. A. Buragga and N. Zaman, "Software development techniques for constructive information systems design," *Softw. Dev. Tech. Constr. Inf. Syst. Des.*, pp. 1–460, 2013, doi: 10.4018/978-1-4666-3679-8.

[7] N. Dissanayake, A. Jayatilaka, M. Zahedi, and M. A. Babar, "Software security patch management - A systematic literature review of challenges, approaches, tools and practices," *Inf. Softw. Technol.*, vol. 144, no. August 2021, p. 106771, 2022, doi: 10.1016/j.infsof.2021.106771.

[8] P. Kruchten, "Contextualizing agile software development," *J. Softw. Evol. Process*, vol. 25, no. 4, pp. 351–361, Apr. 2013, doi: 10.1002/SMR.572.

[9] S. Von Solms and L. A. Futcher, "Adaption of a Secure Software Development Methodology for Secure Engineering Design," *IEEE Access*, vol. 8, pp. 125630–125637, 2020, doi: 10.1109/ACCESS.2020.3007355.

[10] "Handbook of e-Business Security - Google Books." https://books.google.com.sa/books?hl=en&lr=lang_en&id=Zl0PEAAAQBAJ&oi=fnd&pg=PP1&dq=Secure+Software+development+noor+zaman&ots=hsZsRphqlO&sig=mkIjFYi8vVJa7ydDF24tIjqyyqao&redir_esc=y#v=onepage&q=Secure Software development noor zaman&f=false (accessed Aug. 30, 2023).

[11] A. Maria, I. Nazurl, and J. NZ, "A Lightweight and Secure Authentication Scheme for IoT Based E-Health Application," *Int. J. Comput. Sci. Netw. Secur.*, vol. 19, no. 1, pp. 107–120, 2019.

[12] C. Cowan, "Software Security for Open-Source Systems," *IEEE Secur. Priv.*, vol. 1, no. 1, pp. 38–45, Jan. 2003, doi: 10.1109/MSECP.2003.1176994.

[13] S. Kaur, "Security Issues in Open-Source Software," *Int. J. Comput. Sci. Commun.*, vol. 11, no. 2, pp. 47–51, 2020.

[14] "Why tech firms pay hackers to hack them | Hacking | The Guardian." https://www.theguardian.com/technology/2015/nov/14/hackers-technology-bounty-discover-flaws (accessed Aug. 29, 2023).

[15] Z. Liu, Y. Zeng, Y. Yan, P. Zhang, and Y. Wang, "Machine Learning for Analyzing

Malware," *J. Cyber Secur. Mobil.*, pp. 227–244–227–244, Nov. 2017, doi: 10.13052/2245-1439.631.

[16] "Vulnerability in DEC software security kits," *Netw. Secur.*, vol. 1996, no. 7, p. 2, Jul. 1996, doi: 10.1016/s1353-4858(96)90089-6.

[17] T. Le Texier and D. W. Versailles, "Open source software governance serving technological agility: The case of open source software within the DoD," *Int. J. Open Source Softw. Process.*, vol. 1, no. 2, pp. 14–27, 2009, doi: 10.4018/jossp.2009040102.

[18] P. Kamthan, "A perspective on software engineering education with open source software," *Int. J. Open Source Softw. Process.*, vol. 4, no. 3, pp. 13–25, 2012, doi: 10.4018/ijossp.2012070102.

[19] J. W. Njuki, G. M. Muketha, and J. G. Ndia, "A Systematic Literature Review on Security Indicators for Open-Source Enterprise Resource Planning Software," *Int. J. Softw. Eng. Appl.*, vol. 13, no. 3, pp. 27–38, 2022, doi: 10.5121/ijsea.2022.13303.

[20] MarijanDusica and SenSagar, "Good Practices in Aligning Software Engineering Research and Industry Practice," *ACM SIGSOFT Softw. Eng. Notes*, vol. 44, no. 3, pp. 65–67, Nov. 2019, doi: 10.1145/3356773.3356812.

[21] S. H. Kok, A. Abdullah, and N. Z. Jhanjhi, "Early detection of crypto-ransomware using pre-encryption detection algorithm," *J. King Saud Univ. - Comput. Inf. Sci.*, vol. 34, no. 5, pp. 1984–1999, 2022, doi: 10.1016/j.jksuci.2020.06.012.

[22] G. Post and A. Kagan, "Computer security and operating system updates," *Inf. Softw. Technol.*, vol. 45, no. 8, pp. 461–467, Jun. 2003, doi: 10.1016/S0950-5849(03)00016-8.

[23] R. Croft, Y. Xie, and M. A. Babar, "Data Preparation for Software Vulnerability Prediction: A Systematic Literature Review," *IEEE Trans. Softw. Eng.*, vol. 49, no. 3, pp. 1044–1063, Mar. 2023, doi: 10.1109/TSE.2022.3171202.

[24] A. Ekelhart, S. Fenz, G. Goluch, M. Steinkellner, and E. Weippl, "XML security - A comparative literature review," *J. Syst. Softw.*, vol. 81, no. 10, pp. 1715–1724, 2008, doi: 10.1016/j.jss.2007.12.763.

[25] similarly hardship, "The art of software security testing identifying." Accessed: Aug. 30, 2023. [Online]. Available: https://www.academia.edu/40584428/The_art_of_software_security_testing_identifying

[26] J. Geismann and E. Bodden, "A systematic literature review of model-driven security engineering for cyber–physical systems," *J. Syst. Softw.*, vol. 169, p. 110697, 2020, doi: 10.1016/j.jss.2020.110697.

[27] "Risks in Software Testing Techniques A Literature Review," *Int. J. Recent Trends Eng. Res.*, vol. 4, no. 4, pp. 379–384, Apr. 6366, doi: 10.23883/IJRTER.2018.4250.OQ4AI.

[28] A. Takanen, J. DeMott, C. Miller, and A. Kettunen, "Fuzzing for software security testing and quality assurance".

[29] R. H. Campbell, J. Al-Muhtadi, P. Naldurg, G. Sampemane, and M. D. Mickunas, "Software Security — Theories and Systems," *ISSS*, vol. 2609, pp. 1–15, Jun. 2003, doi: 10.1007/3-540-36532-X.

[30] S. Muzafar, N. Z. Jhanjhi, N. A. Khan, and F. Ashfaq, "DDoS Attack Detection Approaches in on Software Defined Network," *14th Int. Conf. Math. Actuar. Sci. Comput. Sci. Stat. MACS 2022*, 2022, doi: 10.1109/MACS56771.2022.10022653.

[31] Z. A. Maher, H. Shaikh, M. S. Khan, A. Arbaaeen, and A. Shah, "Factors Affecting Secure Software Development Practices among Developers-An Investigation," *2018 IEEE 5th Int. Conf. Eng. Technol. Appl. Sci. ICETAS 2018*, Jan. 2019, doi:

10.1109/ICETAS.2018.8629168.
[32] I. M. Y. Woon and A. Kankanhalli, "Investigation of IS professionals' intention to practise secure development of applications," *Int. J. Hum. Comput. Stud.*, vol. 65, no. 1, pp. 29–41, Jan. 2007, doi: 10.1016/J.IJHCS.2006.08.003.
[33] M. Ilyas and S. U. Khan, "Software integration in global software development: Challenges for GSD vendors," *J. Softw. Evol. Process*, vol. 29, no. 8, Aug. 2017, doi: 10.1002/SMR.1875.
[34] A. Avižienis, J. C. Laprie, B. Randell, and C. Landwehr, "Basic concepts and taxonomy of dependable and secure computing," *IEEE Trans. Dependable Secur. Comput.*, vol. 1, no. 1, pp. 11–33, Jan. 2004, doi: 10.1109/TDSC.2004.2.
[35] Shafiq, M., Ashraf, H., Ullah, A., Masud, M., Azeem, M., Jhanjhi, N. Z., & Humayun, M. (2021). Robust Cluster-Based Routing Protocol for IoT-Assisted Smart Devices in WSN. Computers, Materials & Continua, 67(3).
[36] S. Verma, S. Kaur, D. B. Rawat, C. Xi, L. T. Alex and N. Zaman Jhanjhi, "Intelligent Framework Using IoT-Based WSNs for Wildfire Detection," in IEEE Access, vol. 9, pp. 48185-48196, 2021, doi: 10.1109/ACCESS.2021.3060549.
[37] Lim, Marcus, Azween Abdullah, N. Z. Jhanjhi, and Mahadevan Supramaniam. "Hidden link prediction in criminal networks using the deep reinforcement learning technique." Computers 8, no. 1 (2019): 8.
[38] Sennan, S., Somula, R., Luhach, A. K., Deverajan, G. G., Alnumay, W., Jhanjhi, N. Z., ... & Sharma, P. (2021). Energy efficient optimal parent selection based routing protocol for Internet of Things using firefly optimization algorithm. Transactions on Emerging Telecommunications Technologies, 32(8), e4171.
[39] Adeyemo Victor Elijah, Azween Abdullah, NZ JhanJhi, Mahadevan Supramaniam and Balogun Abdullateef O, "Ensemble and Deep-Learning Methods for Two-Class and Multi-Attack Anomaly Intrusion Detection: An Empirical Study" International Journal of Advanced Computer Science and Applications(IJACSA), 10(9), 2019. http://dx.doi.org/10.14569/IJACSA.2019.0100969
[40] Gaur, L., Singh, G., Solanki, A., Jhanjhi, N. Z., Bhatia, U., Sharma, S., ... & Kim, W. (2021). Disposition of youth in predicting sustainable development goals using the neuro-fuzzy and random forest algorithms. Human-Centric Computing and Information Sciences, 11, NA.
[41] Hussain, K., Hussain, S. J., Jhanjhi, N. Z., & Humayun, M. (2019, April). SYN flood attack detection based on bayes estimator (SFADBE) for MANET. In 2019 International Conference on Computer and Information Sciences (ICCIS) (pp. 1-4). IEEE.
[42] Muneer Ahmad, Fawad Ali Khan, and N. Z. Jhanjhi. "An enhanced Predictive heterogeneous ensemble model for breast cancer prediction." Biomedical Signal Processing and Control 72 (2022): 103279.
[43] Gaur, L., Afaq, A., Solanki, A., Singh, G., Sharma, S., Jhanjhi, N. Z., ... & Le, D. N. (2021). Capitalizing on big data and revolutionary 5G technology: Extracting and visualizing ratings and reviews of global chain hotels. Computers and Electrical Engineering, 95, 107374.
[44] Kumar, Tanesh, Bishwajeet Pandey, S. H. A. Mussavi, and Noor Zaman. "CTHS based energy efficient thermal aware image ALU design on FPGA." Wireless Personal Communications 85 (2015): 671-696.
[45] Gouda, W., Sama, N. U., Al-Waakid, G., Humayun, M., & Jhanjhi, N. Z. (2022, June). Detection of skin cancer based on skin lesion images using deep learning.

In Healthcare (Vol. 10, No. 7, p. 1183). MDPI.
[46] C. Diwaker et al., "A New Model for Predicting Component-Based Software Reliability Using Soft Computing," in IEEE Access, vol. 7, pp. 147191-147203, 2019, doi: 10.1109/ACCESS.2019.2946862.
[47] Hussain, S. J., Ahmed, U., Liaquat, H., Mir, S., Jhanjhi, N. Z., & Humayun, M. (2019, April). IMIAD: intelligent malware identification for android platform. In 2019 International Conference on Computer and Information Sciences (ICCIS) (pp. 1-6). IEEE.
[48] Humayun, M., Alsaqer, M. S., & Jhanjhi, N. (2022). Energy optimization for smart cities using iot. Applied Artificial Intelligence, 36(1), 2037255.
[49] M. Lim, A. Abdullah, N. Z. Jhanjhi, M. Khurram Khan and M. Supramaniam, "Link Prediction in Time-Evolving Criminal Network With Deep Reinforcement Learning Technique," in IEEE Access, vol. 7, pp. 184797-184807, 2019, doi: 10.1109/ACCESS.2019.2958873.
[50] Ghosh, G., Verma, S., Jhanjhi, N. Z., & Talib, M. N. (2020, December). Secure surveillance system using chaotic image encryption technique. In IOP conference series: materials science and engineering (Vol. 993, No. 1, p. 012062). IOP Publishing.
[51] Humayun, M., Ashfaq, F., Jhanjhi, N. Z., & Alsadun, M. K. (2022). Traffic management: Multi-scale vehicle detection in varying weather conditions using yolov4 and spatial pyramid pooling network. Electronics, 11(17), 2748.
[52] Shahid, H., Ashraf, H., Javed, H., Humayun, M., Jhanjhi, N. Z., & AlZain, M. A. (2021). Energy optimised security against wormhole attack in iot-based wireless sensor networks. Comput. Mater. Contin, 68(2), 1967-81.
[53] Almusaylim, Z. A., Zaman, N., & Jung, L. T. (2018, August). Proposing a data privacy aware protocol for roadside accident video reporting service using 5G in Vehicular Cloud Networks Environment. In 2018 4th International conference on computer and information sciences (ICCOINS) (pp. 1-5). IEEE.
[54] Wassan, S., Chen, X., Shen, T., Waqar, M., & Jhanjhi, N. Z. (2021). Amazon product sentiment analysis using machine learning techniques. Revista Argentina de Clínica Psicológica, 30(1), 695.
[55] Singhal, V., Jain, S. S., Anand, D., Singh, A., Verma, S., Rodrigues, J. J., ... & Iwendi, C. (2020). Artificial intelligence enabled road vehicle-train collision risk assessment framework for unmanned railway level crossings. IEEE Access, 8, 113790-113806.
[56] Khalil, M. I., Jhanjhi, N. Z., Humayun, M., Sivanesan, S., Masud, M., & Hossain, M. S. (2021). Hybrid smart grid with sustainable energy efficient resources for smart cities. sustainable energy technologies and assessments, 46, 101211.
[57] A. Almusaylim, Z., Jhanjhi, N. Z., & Alhumam, A. (2020). Detection and mitigation of RPL rank and version number attacks in the internet of things: SRPL-RP. Sensors, 20(21), 5997.
[58] Srinivasan, K., Garg, L., Datta, D., Alaboudi, A. A., Jhanjhi, N. Z., Agarwal, R., & Thomas, A. G. (2021). Performance comparison of deep cnn models for detecting driver's distraction. CMC-Computers, Materials & Continua, 68(3), 4109-4124.
[59] Lim, M., Abdullah, A., & Jhanjhi, N. Z. (2021). Performance optimization of criminal network hidden link prediction model with deep reinforcement learning. Journal of King Saud University-Computer and Information Sciences, 33(10), 1202-1210.
[60] Kok, S. H., Azween, A., & Jhanjhi, N. Z. (2020). Evaluation metric for crypto-ransomware detection using machine learning. Journal of Information Security and Applications, 55, 102646.

[61]     Fatima-tuz-Zahra, N. Jhanjhi, S. N. Brohi and N. A. Malik, "Proposing a Rank and Wormhole Attack Detection Framework using Machine Learning," 2019 13th International Conference on Mathematics, Actuarial Science, Computer Science and Statistics (MACS), Karachi, Pakistan, 2019, pp. 1-9, doi: 10.1109/MACS48846.2019.9024821.